# Interscale Mixing Microscopy: far field imaging beyond the diffraction limit

CHRISTOPHER M. ROBERTS,[1,*] NICOLAS OLIVIER,[2] WILLIAM P. WARDLEY,[2] SANDEEP INAMPUDI,[1,&] WAYNE DICKSON,[2] ANATOLY V. ZAYATS,[2] VIKTOR A. PODOLSKIY[1]

[1]*Department of Physics and Applied Physics, University of Massachusetts Lowell, Lowell, MA 01854*
[2]*Department of Physics, King's College London, Strand, London WC2R 2LS, United Kingdom*
[&]*Present address: ECE Department, Northeastern University, Boston, MA, 02115*
*Corresponding author: Christopher_Roberts@student.uml.edu*

**We present an analytical description and an experimental realization of interscale mixing microscopy, a diffraction-based imaging technique that is capable of detecting wavelength/10 objects in far-field measurements with both coherent and incoherent broadband light. This method aims at recovering the spatial spectrum of light diffracted by a subwavelength object based on far-field measurements of the interference created by the object and a finite diffraction grating. A single measurement, analyzing the multiple diffraction orders, is often sufficient to determine the parameters of the object. The presented formalism opens the door for spectroscopy of nanoscale objects in the far-field.**

## 1. Introduction

The detection and visualization of sub-wavelength objects has numerous applications in imaging, spectroscopy, material science, biology, healthcare, and security. In conventional optical microscopy, the fundamental resolution limit is often associated with diffraction criteria by Rayleigh and Abbe[1,2]. According to these criteria, the smallest feature one can optically resolve is approximately limited to ($\lambda_0/2NA$) with $\lambda_0$ being vacuum wavelength and $NA$ being numerical aperture of the objective[3] (Fig.1). In practical terms, however, the position of an isolated object can be determined with much better accuracy; and recent methods relying on the control of the fluorescent state of molecules have used this property to achieve imaging with nanometer resolution[4,5,6]. However, these methods — as well as other optical super-resolution techniques[7] are limited to the study of fluorescent objects. Over the recent years, several label-free optical imaging techniques relying on first-order diffraction and multiple measurements to access high-spatial frequency components of the object have been demonstrated, such as structured illumination microscopy (SIM)[8] or the far field superlens (FSL)[9,27]. When an object with linear photo-response is imaged, the resolution of these techniques is fundamentally limited to approximately $\lambda_0/4$[10]. The imaging of unknown objects with deep sub-wavelength resolution is still predominately performed by relatively slow near-field scanning optical microscopy (NSOM)[11].

There exist several theoretical proposals to achieve fast imaging with sub-wavelength resolution in diffractive and tomographic setups [12,13]. In interscale mixing microscopy (IMM)[14,15], a diffractive element is employed to out-couple information about sub-wavelength features of the object into the far field, similarly to SIM, with the final image formed using a post-processing of the information carried by multiple diffractive orders (Fig.1b). Although IMM somewhat resembles SIM and FSL, in contrast to the latter approaches, IMM utilizes multiple diffractive orders of the grating and therefore provides a resolution that is limited only by experimental noise. Here we report an experimental realization of the IMM with both coherent ($\lambda_0 = 532\ nm$) and incoherent ($600 \leq \lambda_0 \leq 800\ nm$) illumination, achieving a resolution of the order of 70 nm ($\sim\lambda_0/10$) with far-field measurements which do not require point-by-point imaging. Moreover, we present a simple analytical technique to post-process the resulting information, often on the basis of a *single* diffractive measurement. The formalism presented here opens the avenue for far-field microscopy and spectroscopy of nanoscale objects.

Mathematically, the process of imaging is equivalent to recovering the distribution of electromagnetic waves at the location of the object. For one-dimensional objects, this distribution can be described by plane-wave spectrum parameterized by the transverse component of wavevector $k_x$ with wavevector-dependent amplitude $\vec{E}(k_x)$ (2D

objects and combinations of small 3D objects can be considered similarly [15])

$$\vec{E}(\vec{r}) = \int_{-\infty}^{\infty} \vec{E}(k_x) e^{i\vec{k}\cdot\vec{r}-i\omega t} dk_x \quad (1)$$

where $\omega$ is angular frequency, and $\vec{k}$ is the wavevector of the component of the wave. Since the components of the wavevector in free space are related to each other via $k_x^2 + k_z^2 = 4\pi^2/\lambda_0^2$, and since information on the lengthscale $L$ is encoded into wavevectors with $k_x \sim 2\pi/L$, (Figure 1) the information about the sub-wavelength features of the object is encoded in the evanescent, non-propagating waves which exponentially decay away from the source. As the distance to the object increases, the "signal" about the subwavelength features gets exponentially suppressed by the diffraction-limited "background" and experimental noise. Therefore, the resolution of far-field optical microscopy is directly related to the signal-to-noise ratio of the measurements.[16] Sparsity-based imaging [17] fills-in the sub-wavelength information based on analytical continuation of available diffraction-limited measurements of relatively sparse (isolated) objects.

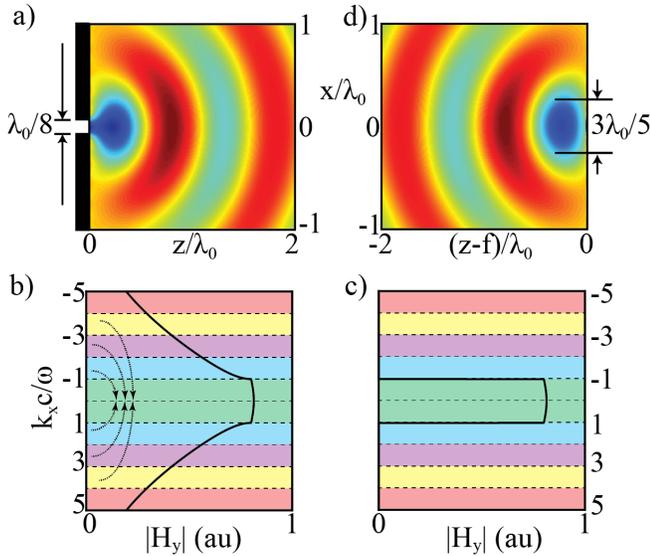

Figure 1: Light scattered by an object (slit in this example) can be represented as a collection of plane waves parameterized by the transverse wavevector $k_x$; The Fourier spectrum of a sub-wavelength object is dominated by evanescent modes with $\frac{k_x c}{\omega} > 1$ (a,b). As the light propagates to the far-field of the object, the evanescent components get exponentially suppressed (c). Attempts to optically re-construct the image in the far-field result in a diffraction-limited pattern (d). IMM mixes evanescent information into the propagating part of the spectrum [arrows in (b)], enabling to distinguish information about sub-wavelength features of the source with far-field measurements.

For densely placed objects, several techniques aim at enhancing the signal carried by evanescent-waves. NSOM collects the near-field information directly, at a cost of slow, point-by-point imaging. The concept of superoscillations [18,19], band-limited functions which oscillate faster than its fastest Fourier component have also been instrumental in achieving sub-wavelength resolution in the scanning microscopy regime. Superlenses [9], hyperlenses[20], and other metamaterial-based solutions have been used to propagate evanescent information in a relatively narrow frequency range. Alternatively, information about $k_x \gg 2\pi/\lambda_0$ can be diffracted into the propagating waves and thus be directly measured in the far field with no need to perform point-by-point scanning.

The use of periodic structures to achieve super-resolution was first investigated a half-century ago by using multiple gratings in order to generate a structured light beam and then using the diffraction of this beam by the object to gain access to the evanescent part of the spectrum. [21,22] Today, diffraction-based characterization allows the analysis of the profile of highly periodic diffracting structures with a resolution of the order of single nm.[23,24] Structured illumination microscopy[25,26] allows the imaging of unknown objects with $\lambda_0/4$ resolution at the expense of multiple measurements and computer post-processing. The far-field superlens[27] uses near field excitation of an object along with resonantly enhanced fields of surface plasmons to achieve $\sim \lambda_0/5$ resolution. Since both SIM and FSL utilize only first order diffraction (for non-fluorescent samples; higher order SIM is only possible on fluorescent samples[28]), these resolution limits cannot be further improved within the existing frameworks.

Several two-dimensional tomography techniques have been proposed theoretically[11]. However, the majority of these techniques requires measurements of both field amplitude and phase in order to reconstruct the image, which is accomplished by fitting measured data to simulations in the Born approximation or an effective medium approach[29]. While possible in theory, such fittings are typically prone to both numerical and experimental noise, especially since optical phase can rarely be measured directly.[30]

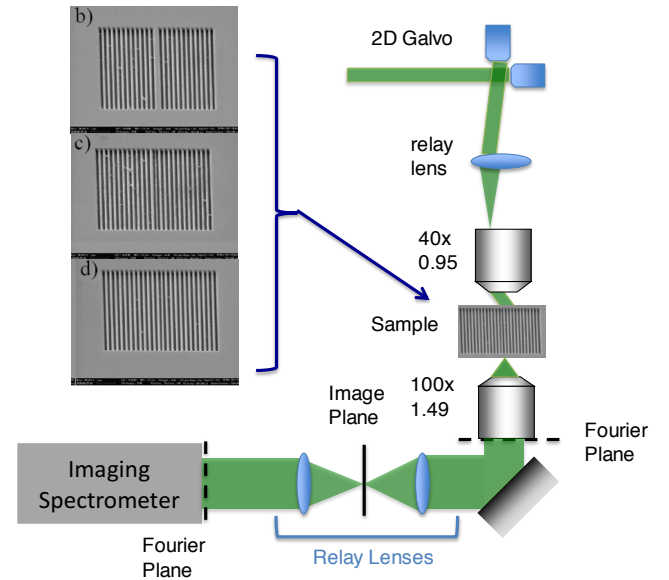

Figure 2 : a) Experimental optical imaging setup; b-d) fabricated gratings with $\Lambda = 275nm$: b) centered defect; c) off-center defect; d) no defects

Interscale mixing microscopy (IMM), first proposed in Ref. [14] relies on a diffraction element — positioned in the near field proximity to an object — to fold the evanescent information into the propagation zone (Figure 1). Such folding essentially mixes the information about sub-wavelength features of the object with the information about diffraction-limited features. Since, in contrast to SIM and FSL, IMM uses multiple diffraction orders, it can in principle achieve unlimited resolution; in addition, as shown here, only a *single measurement* may be sufficient to achieve deep sub-wavelength resolution. In practice, the resolution of IMM is limited by the ability to resolve high-order diffracted signal among the zero-order diffracted background. It has been shown, numerically, that realistic diffracting structures enable up to $\lambda_0/20$ resolution that is limited by experimental noise and instabilities in numerical calculations that can be somewhat mitigated with multiple measurements.[15]

## 2. Experimental Setup

A series of finite gratings (25 periods) were fabricated on a thin (100 nm) gold film on a glass substrate. Two different sets of samples, with periods $\Lambda = 319\ nm$ and $\Lambda = 266\ nm$, each having a slit width of 125 nm were fabricated. Each set contains three structures: (i) a reference, perfectly periodic grating, (ii) a grating with an absent slit in the center, and (iii) a grating with a slit misplaced by ~$\Lambda/4$ (Fig. 2). Apart from the scaling, the results from samples of two different periods are similar to each other. Here we present results for the set of samples with smaller period, where the smallest defect was on the order of ~$70\ nm$. The samples were characterized using a 532 nm CW laser, as well as with a broad-band incoherent white-light illumination 600–800 $nm$, putting the smallest size of the defect in the range of $\lambda_0/7$–$\lambda_0/10$. For the measurements, the samples were illuminated with a 40x objective (0.95 NA) using a quasi-plane wave obtained by focusing the light at the back focal plane of the objective (Fig. 2). The incidence angle was controlled by a pair of galvanometric mirrors conjugated to the focal plane of the illumination objective and the scattered light was detected in transmission using a 100x objective (1.49 NA). The back focal plane of the detection objective (which corresponds to the Fourier plane) was then directly imaged onto an imaging spectrometer. Coherent measurements were performed by imaging the back focal plane directly onto the camera, while white-light measurements were performed with a 300 $\mu m$ wide slit used to select only the specular transmission in the axis perpendicular to the grating, producing far-field diffraction pattern as a function of wavelength.

## 3. Theoretical analysis

Numerical extraction of information about the unknown object was suggested as a basis for IMM in previous studies[14,15]. However, it was shown that in practice the implementation of such algorithms require extreme efforts in equipment calibration[15,24]. Here we present a new, analytical implementation of IMM. The reference diffraction element used in this work represents a finite grating with $N = 25$ slits with width $w$ and period $\Lambda$. The field generated in the near-field proximity of such grating, illuminated by a single, normally incident plane wave, is given by

$$H_y(x) = \sum_{n=-\frac{N-1}{2}}^{\frac{N-1}{2}} rect\left(\frac{x}{w}\right) * \delta(x - n\Lambda) \quad (2)$$

with $*$ denoting the convolution operation, $rect(\xi)$ being the rectangular step function, and $\delta(\xi)$ being the Dirac delta function. The (magnetic) field in the far-field regime is well-approximated by Fraunhofer scalar diffraction theory and is best characterized in the Fourier domain as

$$H_y(k_x) = w\ sinc\left(\frac{k_x w}{2}\right) \frac{sin\left(\frac{Nk_x\Lambda}{2}\right)}{sin\left(\frac{k_x\Lambda}{2}\right)} \quad (3)$$

When an object (e.g., grating defect) of size $l$ is added to the grating, it modifies the field generated by the grating by adding (or subtracting) the field of the object, $H_y^d(x) = rect\left(\frac{x}{l}\right) * \delta(x-a)$. As result, the far-field is modified as well:

$$H_y(k) = w\ sinc\left(\frac{kw}{2}\right) \frac{sin\left(\frac{Nk\Lambda}{2}\right)}{sin\left(\frac{k\Lambda}{2}\right)} + l\ sinc\left(\frac{kl}{2}\right) e^{-ika} \quad (4)$$

The vast majority of optical measurements rely on detecting the intensity, not the field itself. The total intensity measured by the detector is proportional to

$$I(k_x) \propto \left|H_y(k_x)\right|^2 = \frac{I_g(k_x) + I_d(k_x) + I_i(k_x)}{sin^2\frac{k_x\Lambda}{2}} \quad (5)$$

where

$$I_g(k_x) = w^2 sinc^2\left(\frac{k_x w}{2}\right) sin^2\left(\frac{Nk_x\Lambda}{2}\right)$$

$$I_d(k_x) = l^2 sinc^2\left(\frac{k_x l}{2}\right) sin^2\left(\frac{k_x\Lambda}{2}\right) \quad (6)$$

$$I_i(k_x) = 2wl\ sinc\left(\frac{k_x w}{2}\right) sinc\left(\frac{k_x l}{2}\right) sin\left(\frac{k_x\Lambda}{2}\right)$$
$$\times sin\left(\frac{Nk_x\Lambda}{2}\right) cos(k_x a)$$

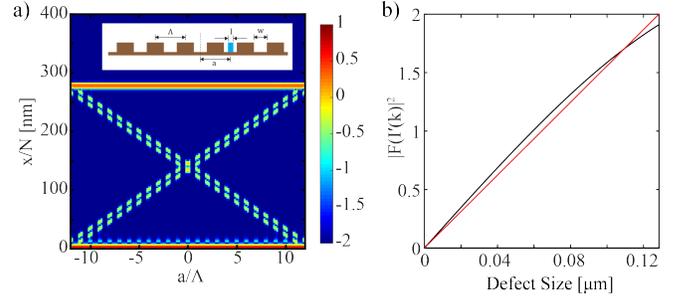

Figure 3 : IMM signal [Eq.(7) for a point object that is being moved from one end to another end of the finite diffraction grating [inset in (a)]: (a) the dependence of the IMM signal on defect position for object size $l = \lambda_0/20$ in $\Lambda = 275nm$; $N = 25$ grating; (b) the dependence of the IMM maximum at $a/\Lambda = 0$ on object size $l$ (black line); red line in (b) shows linear dependence; in both panels $\lambda_0 = 532nm$

The first of these terms represents the spectrum of an ideal grating and dominates the far-field intensity, especially in the proximity of diffraction maxima ($k\Lambda \simeq 2\pi n$). The second term describes the (relatively weak) contribution of the object in the absence of the grating. The last term corresponds to the interference between the fields of the ideal grating and the object. The physics behind the IMM can now be clearly seen from analysis of relative amplitudes of the three terms: while the direct scattered contribution to the signal from the object is small, the grating-object interaction enhances the far field contribution from the object by a factor of $w/l \gg 1$.

The equations above can be used to illustrate the difference between the IMM and SIM-like diffraction techniques. In the IMM formalism, the sample is interrogated by a single plane wave, and signal processing is focused on the cancellation of all principal diffracted beams, analyzing the interference in the ringing tails of the principle orders. As a result, the information about the object can be extracted based on single measurement. In contrast, in SIM-like techniques the sample is interrogated by multiple beams, and the measurement aims at detecting the zero-order interference, which needs to be post-processed and requires multiple measurements.

On the implementation level, the signal-carrying interference term can be enhanced by analyzing the product $I(k_x)\sin^2(k_x\Lambda/2)$ that suppresses the intensity around the diffraction maxima of the ideal grating. The spatial profile and the position of the object can be analyzed by considering the Fourier transform of the power spectrum

$$I(x) = \left|F[I(k_x)\ sin^2\left(\frac{k_x\Lambda}{2}\right)]\right|^2 \quad (7)$$

that formally translates the spectrum from wavevector- into real-space domain.

Figure 3 illustrates the evolution of such spectrum as the defect is moved across the grating. For a fixed position of the defect (inset in Fig. 3a), the signal $I(x)$ has two main features. The first is a maximum located at $N\Lambda$ that corresponds to the period of the grating and represents the contribution from the main diffraction peak. The second feature represents the interaction of the object with the grating. Its position is directly related to the position of the object with a resolution of the order of the slit of the grating. Its intensity, in the limit $l \ll w$, is proportional to the size of the defect (Figure 3). When the object size approaches the size of the grating slit, the intensity dependence on the size slowly deviates from a linear relationship. The signal from the object itself, ($\propto l'^2$ term in Eq.(6)) is not seen in these spectra.

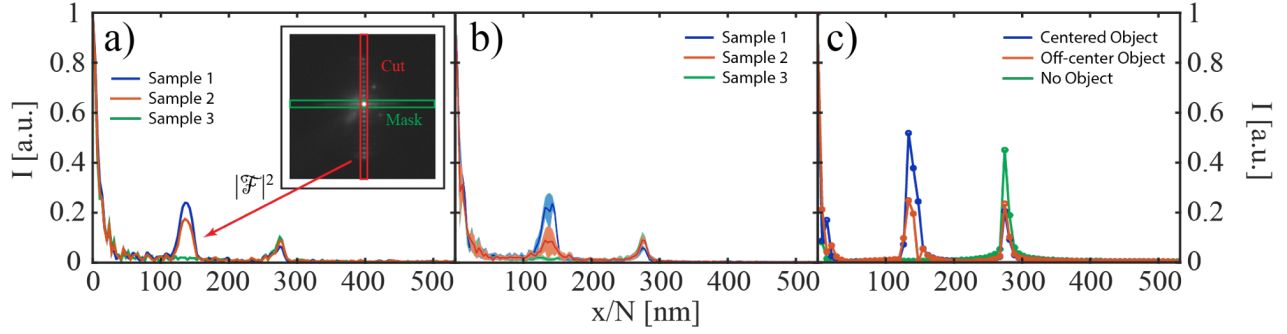

Figure 4: IMM signal based on (a) single experimental measurement corresponding to an incident angle of $21°$, (b) post-processed experimental data of 9 different incident angles, and (c) theoretical predictions of $|\mathcal{F}[I'(k)]|^2$ according to Eq.(7). Thick lines and shaded areas in (b) represent the mean and standard deviation of post-processed data respectively; in all panels $\Lambda = 275 nm, \lambda_0 = 532 nm$. Inset in (a) illustrates the typical imaging processing routine: starting from the raw CCD image, we extract the transmission perpendicular to the grating, suppress the main diffraction peaks by multiplying the signal by $\sin(k\Lambda/2)^2$, followed by Fourier transformation of the power spectrum.

Note that (i) the signatures of sub-wavelength objects can be clearly resolved and (ii) the Fourier spectrum, constructed according to Eq. (5), represents a direct measurement of both position and size of the object with sub-wavelength accuracy.

The power spectrum is dominated by the terms proportional to $\sin\left(\frac{Nk\Lambda}{2}\right)$. The product $I(k_x) \sin^2\left(\frac{k_x\Lambda}{2}\right)$ determines the true resolution of the imaging system. The first term representing the ideal grating will have its centroid located at double the frequency of the interference term due to the square of the dominant sine term. This allows determination of the offset location of the object ($a$) to great precision as demonstrated in Figure 3. The secondary peak exhibits a symmetry about $x = \frac{N\Lambda}{2}$, which leads to positional ambiguity. This ambiguity can be avoided in a non-static arrangement where the imaging target, moving over the periodic aperture, break the $\pm x$ symmetry and provides unambiguous position measurement of the object.

## 4. Results

The diffraction patterns for the structures under consideration, were experimentally measured and simulated using Eq. (7) Figs. 4 and 5. Two different types of measurements were analyzed.

In the first set of measurements, the samples were interrogated with coherent $\lambda = 532$ nm light, for different incident angles. To compare the predictions of theoretical analysis to experimental spectra, raw CCD images were de-skewed to compensate for misalignment of the optical setup (this eliminated a small, $\sim 5/250$ drift of the image [Fig.4a, inset]), and translated to superimpose the images corresponding to different incident angles. In addition to de-skewing and image re-alignment to compensate for off-normal incidence, the contrast of white-light images was enhanced by subtracting the background. The resultant $k_x$ spectra were Fourier-transformed to convert the angular information into the real-space domain according to Eq. (7). Fig. 4a illustrates the result of post-processing a single measurement. Fig. 4b represents the statistics of 9 measurements corresponding to different incident angles. Fig. 4c shows theoretical predictions according to Eq. (7). It is seen that experimental data are in good agreement with theoretical predictions, demonstrating that IMM can resolve $\sim 70$-nm-scale defects. Interestingly, a single measurement provides enough information to resolve the defect.

In the second set of measurements, the sample was illuminated by a broadband incoherent source. The resulting experimental data is compared to theoretical predictions in Fig. 5. Once again, it is seen that the predictions of Eq. (7) are in agreement with experimental data and that both half-period and quarter period defects can be clearly distinguished from the ideal grating and from each other. The spectrum of the defect-free grating has a single wavelength-independent peak representing the period of the grating. The spectra of the two samples with defects contain additional features. The location of these features represent the position of the defect (central slit of the grating), while their structure describes the characteristic of the defect: the missing slit sample yields a double-peak pattern, the shifted slit yields a more complex, broader (but lower amplitude) spectrum maximum. Note that due to the mutual alignment of optical elements and CCDs the broadband spectra provide approximately 4 times the number of $k_x$ datapoints for each wavelength as compared to single-wavelength setup. As result, the fine-structure of the IMM signal can be resolved much better in Fig.5 than in Fig.4.

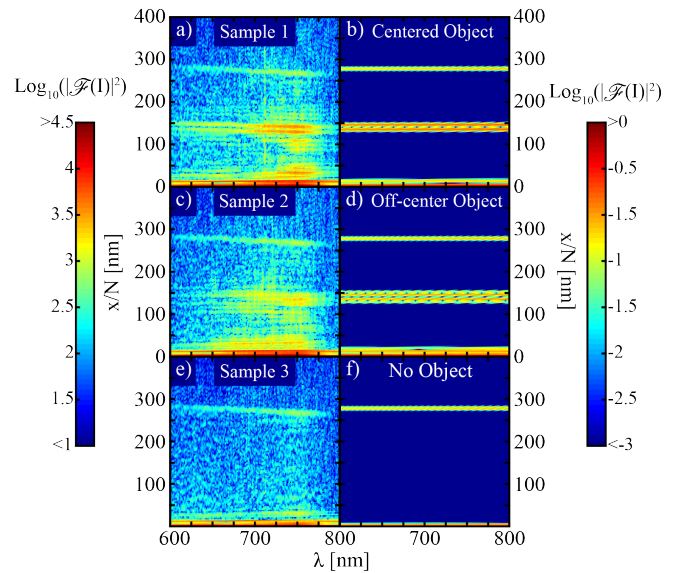

Figure 5: IMM signal extracted from broadband experimental data (a,c,e) and calculated using Eq.(7) (b,d,f) for samples with missing slit (a,b), shifted slit (c,d), and defect-free grating (e,f).

## 5. Discussion and conclusions

We have demonstrated that by using multiple diffractive orders of the grating in a near-field proximity to a nanoscale object, it is possible to recover deep sub-wavelength information, usually lost in the evanescent wave spectrum, in the far-field measurements. A simple, analytical technique capable of predicting the position and size of the opaque object in close proximity to a diffractive element has been developed and validated in the experiments. The resolution is related to recovering the parameters of the full spatial spectrum of light diffracted by the sample and grating system. Notably, the developed formalism has been shown to work for both coherent and incoherent excitation, opening the pathway to the spectroscopy of nanoscale objects. The approach presented here for one-dimensional objects can be extended to the imaging and spectroscopy of two-dimensional structures with the help of 2D diffractive elements.

In realistic objects, the presented formalism will measure the combination of the geometrical size and optical transparency of the object. The spectral dependence of this product, determined by the developed formalism, provides the direct measurement of the spectral response of the nanoscale object.

**Funding** This research was sponsored, in part, by NSF (grants ## ECCS-1102183 and DMR-1209761) and EPSRC (U.K.). A.Z. acknowledges support from the Royal Society and the Wolfson Foundation. The data access statement: all data supporting this research are provided in full in the results section.